\documentclass{aa}
\usepackage{epsfig,times}
\usepackage{color}
\usepackage{graphicx}

\begin{document}
%
%
\title{Magnetically driven superluminal motion from rotating black holes}

\subtitle{Solution of the magnetic wind equation in Kerr metric}

\author{ Christian Fendt \and Jochen Greiner}
\offprints{Christian Fendt, e-mail: cfendt@aip.de }
\institute{Astrophysikalisches Institut Potsdam, An der Sternwarte 16,
D-14482 Potsdam, Germany}

\date{Received <date> ; accepted <date> }

\authorrunning{Fendt \& Greiner}
\titlerunning{General relativistic magnetic jets}


\abstract{
We have investigated magnetically driven superluminal jets originating
from rotating black holes.
The stationary, general relativistic, magnetohydrodynamic wind equation
along collimating magnetic flux surfaces has been solved numerically.
Our jet solutions are calculated on a global scale of a spatial range 
from several to several 1000 gravitational radii.
Different magnetic field geometries were investigated, parameterized by
the shape of the magnetic flux surface and the magnetic flux distribution.
For a given magnetic flux surface we obtain the complete set of physical
parameters for the jet flow.
In particular, we apply our results to the Galactic superluminal sources 
GRS\,1915+105 and GRO\,1655-40.
Motivated by the huge size indicated for the Galactic superluminal knots of
about $10^9$ Schwarzschild radii,
we point out the possibility that the jet collimation process in these sources
may be less efficient and therefore intrinsically different to the AGN.
Our results show that the observed speed of more than 0.9\,c can be
achieved in general by magnetohydrodynamic acceleration.
The velocity distribution along the magnetic field has a saturating
profile. 
The asymptotic jet velocity depends either on the plasma magnetization
(for a fixed field structure) or on the magnetic flux distribution (for
fixed magnetization).
The distance where the asymptotic velocity is reached, is below the
observational resolution for GRS\,1915+105 by several orders of magnitude.
Further, we find that highly relativistic speeds can be reached also for
jets {\em not} emerging from a region close to the black hole, if the flow
magnetization is sufficiently large.
The plasma temperature rapidly decreases from about $10^{10}$K at the foot
point of the jet to about $10^6$K at a distance of $5000$ gravitational 
radii from the source.
Temperature and the mass density follow a power law distribution with the
radius.
The jet magnetic field is dominated by the toroidal component, whereas the
velocity field is dominated by the poloidal component.
%
}

\maketitle
%
\def\etal{{et\,al.}}
\def\mdot{$\dot M$}
\newbox\grsign \setbox\grsign=\hbox{$>$}
\newdimen\grdimen \grdimen=\ht\grsign
\newbox\laxbox \newbox\gaxbox
\setbox\gaxbox=\hbox{\raise.5ex\hbox{$>$}\llap
     {\lower.5ex\hbox{$\sim$}}}\ht1=\grdimen\dp1=0pt
\setbox\laxbox=\hbox{\raise.5ex\hbox{$<$}\llap
     {\lower.5ex\hbox{$\sim$}}}\ht2=\grdimen\dp2=0pt
\def\gax{$\mathrel{\copy\gaxbox}$}
\def\lax{$\mathrel{\copy\laxbox}$}
\def\cm{{\rm cm}}
\def\msun{{\rm M}_{\sun}}
\def\tom{\tilde{\omega}}
\def\toml{\tilde{\omega}_{\rm L}}
\def\omf{\Omega_{\rm F}}
\def\omh{\Omega_{\rm H}}
\def\rl{R_{\rm L}}
\def\rj{R_{\rm jet}}
\def\rh{r_{\rm H}}
\def\rg{r_{\rm g}}
\def\ra{R_{\rm A}}
\def\xa{x_{\rm A}}
\def\gi{g_{\rm I}}
\def\cs{c_{\rm s}}
\def\xss{x_{\star}}
\def\css{c_{\rm s\star}}
\def\rss{R_{\rm \star}}
\def\nss{n_{\rm \star}}
\def\sigs{\sigma_{\star}}
\def\ups{u_{\rm p\star}}
\def\bps{B_{\rm p\star}}
\def\phs{\Phi_{\star}}
\def\up{u_{\rm p}}
\def\vp{v_{\rm p}}
\def\bp{B_{\rm p}}
\def\tbp{\tilde{B}_{\rm p}}
\def\sm{\sigma_{\rm m}}
\def\aL{\tilde{L}}
\def\grs{GRS\,1915+105\,\,}
\def\gro{GRO\,1655-40\,\,}
\section{Introduction}

\subsection{Relativistic jets and Galactic superluminal motion}
%
Apparent superluminal jet motion originating in the close environment
of a rotating black hole
is observationally indicated for two classes of sources concerning mass
and energy output.
One class is the family of radio loud active galactic nuclei 
(hereafter AGN). 
In the AGN standard model highly relativistic jet motion is
explained by {\em magnetohydrodynamic} processes in a black hole -
accretion disk environment  (for a review see Blandford 1990).
Jets are magnetically accelerated and possibly also collimated by
magnetic forces.
However, the detailed interaction process of the magnetized black
hole - accretion disk system which is believed to lead to 
the ejection of high velocity blobs is not yet fully understood.

The other class are galactic binary systems for which 
radio observations have also detected superluminal motion (see reviews
of Fender 2000 or Greiner 2000).
The two most prominent examples are the high energy sources \grs
(Mirabel \& Rodriguez 1994) 
and \gro (Hjellming \& Rupen 1995; Tingay et al. 1995).
The de-projected jet speed of both sources is $\ga 0.9\,c$ and 
surprisingly similar,
although for \grs also a higher velocity component has been observed
recently (Fender \etal\ 1999).
\gro is a binary consisting of a $7.02\pm 0.22\,\msun$ black hole and 
a $2.3\,\msun$ F-subgiant (Orosz \& Bailyn 1997) at a distance of 3 kpc.
GRS 1915+105 is at 10--12 kpc distance (Fender \etal\ 1999), but the 
component masses of the presumed binary are not known. Order of magnitude
estimates based on X-ray variability and QPO properties range from
10--80 $\msun$ (Morgan \etal\ 1997, Greiner \etal\ 1998).
As for the AGN jet sources, observational evidence for a black 
hole - accretion disk system is found also for the Galactic superluminal 
sources.
Observations have also indicated that accretion disk instabilities 
may be related to jet ejection (Greiner et al. 1996, Belloni et al. 1997,
Mirabel \etal\ 1998).
Therefore, the jet formation process for extragalactic jets and their
Galactic counterparts may be the same, although
the mechanism that accelerates and collimates the \grs ejecta
is yet unclear (Rodriguez \& Mirabel 1999).

Optical polarization measurements have been obtained for the microquasar 
GRO J1655-40 (Scaltriti et al. 1997, Gliozzi et al. 1998). The polarization
angle is approximately parallel to the accretion disk plane. The amount of
polarization has been found to vary smoothly with the orbital phase,
being smallest at binary phase 0.7--0.8. It has been noted that the 
occasionally observed X-ray dips occur at the same phase interval
(Ueda et al. 1998, Kuulkers et al. 1998)
suggesting that it may be related to either a 
thickening of the disk rim at the impact site of the accretion stream
from the companion or the overflow of this stream above/below the disk.
The orbital polarization modulation rules out a synchrotron origin in the jet,
and implies the presence of electron scattering plasma above the accretion 
disk which is asymmetrically distributed or asymmetrically illuminated.
The existence of such scattering plasma is consistent with the interpretation 
of the iron features as observed with ASCA as absorption lines and edges in a 
thick, cool torus of column 
$N_{\rm H} > 10^{23}{\rm cm}^{-2}$ (Ueda et al. 1998).

The relativistic speed observed for the Galactic superluminal sources 
($\sim 0.9-0.98\,c$ de-projected)
corresponds to a bulk Lorentz factor of $\gamma = 2-5$
although this number is not very accurate (e.g. Fender \etal\ 1999).
Therefore, for any theoretical investigation of these objects at least 
special relativity has to be taken into account.
If the superluminal motion originates close to a black hole,
also general relativistic effects may become important.

The ejection of matter itself is not a stationary process.
In \grs also repeated emission of knots is observed
(Rodriguez \& Mirabel 1999). 
X-ray and radio observations suggest that a wide range of ejected mass and 
ejection frequency is possible. 

Though the galactic jet sources are nearby, they are not better resolved
spatially because the distance ratio between AGN and microquasars is 
smaller than their mass ratios.
Nevertheless, an important implication may also come from the observed size
of the superluminal knots which are observationally resolved.
In the case of \grs the characteristic dimension of the 'jet' is 
35\,mas, equivalent to $7\times 10^{15}$\,cm at a distance of 12.5\,kpc 
(Rodriguez \& Mirabel 1999).
We emphasize that such a knot size corresponds to $\sim 10^9$
Schwarzschild radii for $R_{\rm S} = 1.5\times 10^6\,(M/5\,\msun)\cm$!
This is a huge factor and may be in distinctive difference to the
AGN jets.
Similarly, the VLBA data show the core as a collimated jet down to a 
distance of 10\,AU from the central source with an opening
angle of $<10\degr$ (see Mirabel \& Rodriguez 1999)
corresponding to $10^7\,(M/5\msun)$ Schwarzschild radii.
The length of the radio jet is about 100\,AU.

However, when interpreting the observed emission region, one has
to keep in mind that this region may not represent the jet flow 
itself, but some {\em part} of another, larger, structure.
For example, in some extragalactic jet sources there is indication
that the knots travel along helical trajectories,
believed to be prescribed by a large-scale
helical magnetic field of an almost cylindrically collimated jet 
(Zensus et al. 1995; Camenzind \& Krockenberger 1992).

In \gro the motion of the radio knots is complicated and requires (at least)
precession between different ejections (Hjellming \& Rupen 1995). 
The knot structures in \grs remained fixed implying that the whole
knot moves with the same speed without spatial diffusion and with
an axial velocity profile more or less constant.
 
Based on minimum energy arguments and only relativistic electrons
responsible for the synchrotron radiation in the knots of
\grs, Rodriguez \& Mirabel (1999) derive a magnetic
field strength of about 50\,mG to 7\,mG, the decrease resulting
from the expansion of the knot.
They also estimate the rest mass of a knot of $\ge 10^{23}$\,g,
and together with (steady) photon luminosity of
$\simeq 3\times 10^{38}\,{\rm erg\,s^{-1}}$,
exclude radiation as driving mechanism for the knots.

\subsection{Theory of magnetic jets} 
%
From the introductory remarks it is clear that a quantitative
analysis of superluminal motion must take into account both
magnetohydrodynamics (hereafter MHD) and (general) relativity.

The first theoretical formulation of the electromagnetic
force-equilibrium in Kerr space-time around rotating black holes
was given by Blandford \& Znajek (1977) and Znajek (1977),
who discovered the possibility of extracting rotational energy and
angular momentum from the black hole electromagnetically.

Camenzind (1986, 1987) formulated a fully relativistic stationary
description of MHD flows, basically applicable to any field geometry.
The structure of such collimating jet magnetospheres in the case of
Kerr space time was presented by Fendt (1997).
Solutions of the so-called wind equation in Kerr geometry (see below)
considering the stationary plasma motion along the magnetic field
were obtained by Takahashi et al. (1990), however, 
mainly discussing the accretion flow onto the black hole.

While the asymptotic structure of the propagating jets becomes more and
more understood with the help of time-dependent magnetohydrodynamical,
also relativistic, simulations
(e.g. Nishikawa et al. 1997;
Mioduszewski et al. 1997;
Hardee et al. 1998),
the process of jet formation itself and the collimation of the outflow
region is a task still too complex for numerical simulations.
The involved length scales and gradients require a high resolution in
grid size and time stepping.
Koide et al. (1998) were first to perform general relativistic 
MHD simulations of jet formation close to the black hole.
In their model, the interaction of an initially cylindrical magnetic
field with a Keplerian accretion disk results first in an inflow of
matter towards the black hole.
This accretion stream interacts with the hydrostatic corona around 
the black hole giving rise to a relativistic gas pressure driven jet.
At larger radii a magnetically driven wind is initiated from the
accretion disk.
The simulations were performed for less than two rotations of the inner
disk 
(corresponding to less than 0.02 rotations of the disk at the outer edge
of the grid).
Although these results of the first fully general relativistic MHD
simulations look indeed very exciting, some objections can be raised
about the underlying model.
The initial condition applied is that of a hydrostatic corona around
a black hole,
an assumption which is not compatible with the boundary of a 
black hole  horizon.
Such a configuration is not stable and will immediately collapse.
Recently, the authors extended their work applying an initial coronal
structure in steady infall surrounding a non-rotating black hole 
(Koide et al. 1999).
They find a two-layered jet consisting of a magnetically driven jet
around a gas-pressure driven jet.
In addition, Koide et al. (2000) considered the quasi-steady infall 
of the corona around a Kerr black hole.
They find that jet formation seems to differ for co-rotating and 
counter-rotating disks.
The jet ejection tends to be easier in the latter case with a jet
origin much closer to the hole.
Also, a new feature of another magnetically driven (though
sub-relativistic) jet appears within the gas-pressure driven jet.
The computations were lasting over a few inner disk orbits.
Therefore, the observed events of mass ejection could still be a relict
of the initial condition and may not be present in the long-term
evolution.
Clearly, it would be interesting to perform the Koide et al. simulations
for a longer time and look 
whether the mass ejection continues over many disk orbits, 
whether the simulation evolves into a final stationary state 
(as e.g. in Ouyed \& Pudritz 1997, Fendt \& Elstner 2000),
or whether the jet formation retains its unsteady behavior which could 
explain the emission of superluminal knots observed in the relativistic
jets.

\subsection{Aim of the present study}
%
In this paper, a stationary magnetic jet flow along a given magnetic
flux surface is investigated in the context of general relativity.
Due to the stationary approach, we cannot treat any time-dependent
phenomena.
Our emphasis is to trace the {\em large scale} behavior of the flow
from it's origin close to the black hole to large distances.
This is an essential point in particular for the Galactic superluminal
sources because of the possible huge spatial extension of the jets 
compared to the central black hole.
The stationary model allows for a {\em global} treatment of the jet 
flow, 
i.e. an investigation over a large range of magnitudes for density 
and magnetic field strength.
This is not yet feasible with time-dependent MHD codes presently available.
In particular, we address the following topics.

\begin{itemize}
\item
For a given geometry of the magnetic field, which are the resulting jet 
{\em dynamical parameters} as velocity, density or temperature?
\item
How important are the effects of general relativity? 
Does the superluminal flow indeed originate very close to a black hole?
\item
From the investigation of different field geometries we expect some hints
to the jet opening angle and the length scale of the collimation process.
\end{itemize}
The structure of this paper is as follows. 
In Sect.\,2, basic equations for relativistic magnetospheres are reviewed
in the context of Kerr metrics.
In Sect.\,3, the model underlying our numerical calculations is discussed. 
We present our numerical results in Sect.\,4 and discuss solutions
with different geometry and jet parameters. We summarize our paper
in Sect.\,5.

\section{Description of a MHD flow in Kerr metric}
%
Under the assumptions of axisymmetry, stationarity and infinite conductivity,
the MHD equations reduce to a set of two basic equations describing
the local force-balance across the field and along the field 
(for references, see, e.g., Blandford \& Znajek 1977; 
Thorne et al. 1986;
Camenzind 1986, 1987;
Okamoto 1992;
Beskin \& Pariev 1993, Beskin 1997).

The trans-field or {\em Grad-Shafranov equation} determines the field
structure, 
whereas the {\em wind equation} describes the flow dynamics along the
field.
Due to the stationarity assumption, certain conservation laws apply.
The total energy density, the total angular momentum density, 
the mass flow rate per flux surface and the iso-rotation parameter
are conserved quantities along the surfaces of constant magnetic
flux (Camenzind 1986). 

In this paper the motion of a magnetized plasma is calculated from the
wind equation.
The plasma moves along a prescribed axisymmetric magnetic flux surface
which originates in a region close to a rotating black hole.

\subsection{Space-time around rotating black holes}
%
The space-time around a rotating black hole with a mass $M$ and angular
momentum per unit mass $a$ is described using Boyer-Lindquist 
coordinates with the line element
\begin{equation} 
ds^2 = \alpha^2dt^2 - \tom^2\,(d\phi -\omega dt)^2
-(\rho^2/\Delta)\,dr^2 - \rho^2\,d\theta^2,
\end{equation}
where $t$ denotes the global time, 
$\phi$ the angle around the axis of symmetry,
$r,\theta$ similar to there flat space counterpart spherical
coordinates,
and where geometrical units $c=G=1$ have been applied
(see Appendix A for further definitions).
The horizon of the Kerr black hole is located at 
$\rh = M + \sqrt{M^2 - a^2}$.
We will normalize all radii to gravitational radii
$\rg = \rh (a=M) = M$.
The angular velocity of an observer moving with zero angular momentum 
(ZAMO) is $\omega = (d\phi/dt)_{\rm ZAMO}$,
corresponding to the angular velocity of the differentially rotating space.
The lapse function is $\alpha= (d\tau/dt)_{\rm ZAMO}$ describing the lapse
of the proper time $\tau$ in the ZAMO system to the global time $t$.

\subsection{Description of the electromagnetic field}
%
In the 3+1 split of Kerr space time (Thorne et al. 1986) 
the electromagnetic field 
$\vec{B}, \vec{E}$, the current density $\vec{j}$, and the electric
charge density $\rho_{\rm c}$ can be described very similar
to the usual expressions,
if measured by the ZAMO's according to the locally flat
Minkowski space.
These local experiments then have to be put together by a global observer
for a certain global time using the lapse and shift function for the
transformation from the local to the global frame.

With the assumption of axisymmetry a magnetic flux surface can be
defined measuring the magnetic flux through a loop of the Killing
vector $\vec{m} = \tom^2\nabla\phi$,
\begin{equation} 
\Psi (r,\theta) = \frac {1}{2 \pi} \int {\vec {B}}_{\rm p} \cdot
d{\vec{A}}\,,
\quad\quad {\vec{B}}_{\rm p} = \frac{1}{\tom^2}\nabla \Psi \wedge
{\vec{m}},
\end{equation}
corresponding to the magnetic flux through an area $\pi (r\,\sin\theta)^2$ 
around the symmetry axis (in the limit of Minkowski space).

With the assumption of a degenerated magnetosphere, 
$ |\,|\vec{B}|^2 - |\vec{E}|^2 | >> | \vec{E} \cdot \vec{B} | \simeq 0 $
an 'angular velocity of field lines' can be derived from the derivative
of the time component of the vector potential 
$\omf = \omf(\Psi) = -2 \pi c (dA_0/d\Psi) $.
We will denote this quantity with the term `iso-rotation parameter'.

\subsection{The wind equation}

It has been shown that a stationary, polytropic, general relativistic MHD
flow along an axisymmetric flux surface $\Psi(r,\theta)$ can be described
by the following {\em wind equation} for the poloidal velocity 
$\up \equiv \gamma \vp/c$,
\begin{equation} 
u_{\rm p}^2 + 1  = -\sm \left(\frac{E}{\mu}\right)^2
\frac {k_0 k_2  + \sm 2 k_2 M^2 - k_4 M^4}{(k_0 + \sm M^2)^2}\,,
\end{equation}
where
\begin{eqnarray} 
k_0 & = & g_{33} \omf^2 + 2 g_{03} \omf + g_{00},
\nonumber \\
k_2 & = &  1 - \omf \frac{L}{E},
\nonumber \\
k_4 & = & 
- \left(g_{33} + 2 g_{03} \frac{L}{E} + g_{00} \frac{L^2}{E^2} \right) /
         \left( g_{03}^2 - g_{00} g_{33} \right)
\nonumber 
\end{eqnarray}
(Camenzind 1986, Takahashi et al. 1990).
The Alfv\'en Mach number $M$ is defined as $M^2 = 4 \mu n \up^2 / \tbp^2 $,
with the proper particle density $n$, the specific enthalpy $\mu$,
and a poloidal magnetic field $\tbp = \bp / (g_{00)}+g_{03}\omf) $,
rescaled for mathematical convenience.
The quantity $\sm$ stands for the sign of the metric (we have chosen
$\sm = -1$, see appendix A).
For a polytropic gas law with the index $\Gamma \equiv n/m$, 
the wind equation (3) can be converted into a polynomial equation,
\begin{equation} 
\sum\limits_{i=0}^{2n+2m} A_i (x;\Psi,\Phi;\omf;E,L,\sigs)\,
u_{\rm P}^{i/m} = 0 \,,
\end{equation}
(Camenzind 1987, Englmaier 1993, Jensen 1997),
where the coefficients $A_i$ are now defined as functions of the 
normalized cylindrical radius $x=R/\rg$ (see Appendix B).
The shape of the axisymmetric magnetic flux surface $\Psi $ is 
prescribed as function $z(x;\Psi)$.
The flux function $\Phi = \sqrt{-g}\tbp$ describes the opening of the
flux tube. 
The faster $\Phi$ decreases the faster magnetic energy is converted
into kinetic energy.
We define the dimensionless magnetization parameter 
\footnote{ 
Note that this definition for the magnetization varies from the original
Michel magnetization parameter 
$\sigma_{\rm M} = \Phi_{\rm M}^2 / 4\pi f_{\rm M} c \rl^2$, 
where $\Phi_{\rm M}$ is the magnetic flux, $f_{\rm M}$ the mass flux
and $\rl$ the light cylinder. 
Usually, the general relativistic equations are normalized to the
gravitational radius, 
whereas the special relativistic equations are normalized to the light
cylinder}
at the `injection' point $\xss$ following Takahashi et al. (1990),
\begin{equation}
\sigma_{\star} = \frac{\Phi^2_{\star}}{4\pi m_{\rm p} I_{{\rm p}\star}},
\end{equation}
measuring the Poynting flux in terms of particle flux
$I_{\rm p} \equiv \sqrt{-g}\,n\up $, 
where $m_{\rm p}$ is the particle mass (here the proton mass).
The magnetization determines the maximum energy available for plasma
acceleration and thus determines also the asymptotic poloidal 
velocity.
The other wind parameters are total energy density $E$, 
total angular momentum $L$,
and the iso-rotation parameter $\omf$.
The non relativistic limit of Eq.\,(4) has been solved numerically by
Kudoh \& Shibata (1995, 1997).

We choose the polytropic index $\Gamma = 5/3$
for a hot relativistic proton-electron plasma
(a hot electron-positron plasma would imply $\Gamma = 4/3$).
Then, at each radius $x$ the polynomial equation (4) has 
$2n+2m = 16$ solutions. 
Some of these mathematical solutions have no physical meaning,
e.g. because $\up^2$ is negative.
The remaining physical solutions form a bunch of different curves in the 
$\up(x)$-diagram representing different solution branches
(see our solution S1 in Appendix C, Fig.\,C1). 
The unique branch of the 'wind' solution starts at a small radius with 
small velocity continuing outwards with increasing velocity.
For an other parameter choice also 'accretion' branches can be found, 
starting from a large radius with small velocity and continuing inwards
with increasing velocity (not shown in Fig.\,C1).

However, not for all parameters $E,L,\sigma$ there exist
{\em physical} solutions which are continuous functions of $x$ and
therefore defined along the whole flux surface.
It is well known that at the magnetosonic points the wind equation (3)
becomes singular (see Camenzind 1986, Takahashi et al. 1990).
Regularity of the solution requires a flow velocity equal to 
the speed of the MHD waves in order to obtain a smooth 
(self-consistent) transition at the magnetosonic points.
In order to match astrophysical boundary conditions we fix the
following parameters,
\begin{itemize}
\item{} {\em the 'injection' radius, $\xss$}, the location where the matter
couples to the magnetic field. 
This radius also determines the iso-rotation parameter $\omf$.

\item{}
{\em the 'injection' velocity $\ups = \up(\xss)$}, defining the
initial kinetic energy.

\item{}
{\em the Alfv\'en radius $\xa $}, which fixes the total angular momentum
of the flow.
\end{itemize}
The critical wind solution for a given flux surface can then be found by
varying the flow parameters in Eq.\,(4).
Due to numerical convenience, we vary 
\begin{itemize}
\item{}
{\em the sound speed $\css $} at the injection radius, defining the
initial density (or gas pressure and temperature),
\item{}
{\em the magnetization} parameter at the injection point
$\sigs (\Psi) = \phs^2/(4\pi m_{\rm p} I_{{\rm p}\star})$.
\end{itemize}
In turn, the condition of a regular flow at the magnetosonic points 
fixes the sound speed and magnetization and, 
thus, jet mass flow rate and temperature.

\section{The model assumptions}

\subsection{The model in general}
%
Observationally the jet phenomenon of AGN, young stellar objects and 
microquasars is always connected to the signatures of an accretion disk.
We therefore assume a similar disk-jet scenario for the jet formation in
{\em Galactic} superluminal jet sources.
In general our model geometry follows the standard model for jet
formation in AGN 
(cf. Blandford 1990).

Two typical length scales enter the problem.
(i) The gravitational radius $\rg$ measures the influence of gravity on
the metric.
(ii) The asymptotic light cylinder $\rl$ describes the influence of rotation
on the electrodynamics.

\setlength{\unitlength}{1mm}
\begin{figure}
\parbox{60mm}
\thicklines
\epsfysize=70mm
\caption
{Model geometry applied for our numerical calculations. 
The poloidal field structure is prescribed as magnetic flux surfaces
with different opening angle. 
The flux surfaces have different foot point radii along the accretion
disk (not visible).
The central source is a black hole implying that general relativistic
effects have to be taken into account.
The toroidal field follows from the solution of the wind equation.
}
\end{figure}

\subsection{The central black hole}
%
The black hole mass and angular momentum determine the geometry of space.
Since we use dimensionless equations normalized to the gravitational
radius, our results scale with the mass of the black hole. 
For parameter estimates we assume a black hole mass of $5\,\msun$ which is
about the value inferred for the galactic superluminal sources.
The angular momentum $a$ as the other black hole parameter is not known
for any of the relativistic jet sources.
Interpretation of the high effective temperatures of the accretion disk
as well as the stable QPO frequency (as Thirring-Lense effect)
suggests that $a$ \gax 0.9 for \grs and \gro (Zhang \etal\ 1997).
Theoretically, one may expect a rapidly rotating black hole because
of angular momentum conservation during the collapse and also
accretion of angular momentum from the accretion disk (King \& Kolb 2000).
Here, we choose $a = 0.8$, a value which is not extreme, but clearly
different to Schwarzschild metric.
The rotation rate of the black hole is defined as
$\omh \equiv \omega(\rh) = a/(2M\rh)$.
The Kerr parameter $a$ does not influence the solution of the wind 
equation directly.
However, for rotating black holes the marginally stable orbit 
$r_{\rm ms}$ is
closer to the horizon,
$r_{\rm ms}=6\rg $ for $a=0$ and $r_{\rm ms}\simeq \rg$ for $a\simeq 1$
(This is the case for a co-rotating disk. For a retrograde disk rotation
$r_{\rm ms}\simeq 9\rg $ for $a\simeq 1$).
Therefore, assuming that the jet magnetic field is anchored just at the
marginally stable orbit, 
for a rapidly rotating black hole the maximum angular velocity of the jet
foot points increases by a factor of 
$6^{3/2}/2=7.4$.
Correspondingly, the light cylinder radius of the jet moves inward by
the same factor.

In addition to the well-known special relativistic light cylinder,
the differential rotation of the space $\omega$ leads to the formation
of a second light surface.
At this position the 'rotational velocity' of the field lines relative to 
the ZAMO equals the speed of light (see Blandford \& Znajek 1977).
The position of the two light surfaces $\toml$ is defined by 
$\toml^2 = (\pm \alpha \,c / (\omf - \omega))^2$, where
the $+$ ($-$) sign holds for the outer (inner) light surface with 
$\omf > \omega$ ($\omf < \omega$).
However, these light surfaces have no direct implication for the MHD flow.
In the limit of a strong magnetization, the MHD Alfv\'en surfaces
(for inflow and outflow) approach the corresponding light surfaces.

\subsection{The accretion disk}
%
X-ray observations of \grs detected strong intensity variations indicating
major instabilities of an accretion disk (Greiner et al. 1996).
Belloni et al. (1997) find that the highly variable X-ray spectra could be
explained if the inner disk is alternatively removed and replenished due to
a thermal-viscous instability.
Simultaneous X-ray and infrared observations of \grs revealed evidence
for a disk--jet interrelation (Eikenberry et al. 1998, Mirabel \etal\ 1998). 
The observed flares in the X-ray and IR bands have a consistent offset delay
of $\sim 30$\,min indicating an origin from the same event.

The accretion rate in GRS 1915+105 and GRO J1655-40 can be determined 
from the observed X-ray luminosities (e.g. Greiner \etal\ 1998).
Depending on the chosen efficiency (5\% in non-rotating versus 42\%
in maximally rotating black holes) the accretion rate ranges between
$1-9 \times 10^{-7} \msun {\rm yr}^{-1}$ (GRS 1915+105) and
$0.8-7 \times 10^{-8} \msun {\rm yr}^{-1}$ (GRO J1655-40), respectively.

From the theoretical point of view an accretion disk surrounding the black
hole is the essential component concerning magnetic jet formation.
It is considered to be responsible for the following necessary ingredients
for jet formation, propagation, and collimation.

\begin{itemize}
\item{} 
The {\em generation of the magnetic field}. 
In contrast to stellar jets the magnetic field of jets from black holes
cannot be supplied by the central object but has to be generated by the
surrounding accretion disk.  
Dynamo action in general relativistic accretion disks were discussed by
Khanna \& Camenzind (1996a, 1996b) and Brandenburg (1996).

\item{} 
The {\em mass loading of the jet}.
The accreting material becomes partly diverted into the jet. 
Evidently, no mass outflow is possible from the black hole itself,
in difference to a stellar wind.
The (non-relativistic) self-similar accretion-ejection mechanism was 
investigated by Ferreira (1997).

\item{} 
The {\em electric current} system. 
Differential rotation of the disk is also responsible for driving the
poloidal electric current system in the jet magnetosphere. 
Such a current extracts angular momentum from the disk and eventually
allows for mass accretion into the central object.

\end{itemize}

\subsection{Model parameters for the wind motion}
%
\subsubsection{The magnetization parameter}
%
The leading parameter for the wind solution along a fixed poloidal field 
is the magnetization parameter (5).
Re-normalization to astrophysical units gives
\begin{equation} 
\sigs (\Psi) = \frac{\phs^2}{4\pi m_{\rm p} I_{{\rm p}\star} }
 \rightarrow  \frac{\bps^2\rss^4}{c \dot{M}_{\rm jet}(\Psi) \rg^2 }
 = \frac{\bps^2 \rg^2}{c \dot{M}_{\rm jet}(\Psi)}
 \!\!\left(\frac{\rss}{\rg}\right)^{\!\!4}
\end{equation}
where 
$\dot{M}_{\rm jet}(\Psi) 
\simeq 4\pi m_{\rm p} \nss c \ups \rss^2$
is the jet mass flux enclosed by an area
of radius $\rss$.
A first order estimate of the magnetization can be derived from
the disk equipartition field strength.
Then, with a reasonable assumption on the jet mass flow rate related 
to the disk accretion rate, this gives the jet magnetization.
Although the equipartition field strength is model-dependent,
the different models 
(e.g. either advection dominated disk or standard disk,
either Kramer's opacity or Thomson scattering) 
give rather similar results.
A self-similar advection dominated disk model with the accretion rate
$\dot{M}_{\rm acc}$ gives 
\begin{equation}
B_{\rm eq} \simeq 2.5\times 10^9 {\rm G}\,\,\alpha_{\rm vis}^{-\frac{1}{2}} 
\!\left(\frac{M}{5\msun}\right)^{\!\!-\frac{1}{2}} 
\!\left(\frac{\dot{M}_{\rm acc}}{\dot{M}_{\rm E}}\right)^{\!\!\frac{1}{2}} 
\!\left(\frac{R}{\rg}\right)^{\!\!-\frac{5}{4}}\!\!\!\!\!,
\end{equation}
where $\dot{M}_{\rm E} = 1.1\times10^{-7}(M/5\msun)\,\msun {\rm yr^{-1}}$
is the Eddington luminosity and $\alpha_{\rm vis}$ is the viscosity parameter
(see e.g. Narayan et al. 1998).
%
%
In comparison, an optically thin standard accretion disk with Thomson
opacity gives
$
B_{\rm eq} \simeq \sqrt{8\pi P} = \sqrt{8\pi a T^4/3} 
\simeq 1.8\times 10^8 {\rm G}\,\alpha_{\rm vis}^{-1/2}(M/5\msun)^{-1/2} 
(R/\rg)^{-3/4}
$
(see Blandford 1990).
Note that these estimates are only valid within certain limits
of the accretion rate and the disk radius.
Using the advection dominated disk model equipartition field strength,
we obtain the following estimate for the magnetization at the
injection radius,
\begin{equation} 
\sigs (\Psi) = 
16
\frac{1}{\alpha_{\rm vis}} 
\left(\frac{M}{5\msun}\right)
\left(\frac{\dot{M}_{\rm jet}}{\dot{M}_{\rm acc} }\right)^{-1}
\left(\frac{\rss}{\rg}\right)^{3/2} \nonumber \\
\end{equation} 
A comparison with the original Michel magnetization parameter
$\sigma_{\rm M}$
must take into account a factor $(\rg /\rl)^2$.
The magnetization parameter derived from the field distribution in a
standard accretion disk model (see above) will give a similar result.
We emphasize that we do not `apply' a certain disk model (e.g. the ADAF
model) in our computations.
However, a comparison in the context of accretion disk theory just puts
our wind parameters on a safer ground.
Note, that neither the ADAF model nor the standard disk model takes
into account the influence of magnetic fields.
Moreover, the ADAF estimates as cited in Eq.\,(7) rely on the
self-similar assumption.
Compared to the standard disk, by definition, the matter in the ADAF disk
would be rapidly advected possibly influencing also the wind ejection.
However, such a detailed treatment is beyond the scope of this paper
and may only be considered in numerical simulations investigating the
disk-jet interaction itself (Koide et al. 1998, 1999, 2000)

\setlength{\unitlength}{1mm}
\begin{figure}
\parbox{60mm}
\thicklines
\epsfysize=80mm
\caption
{Projected magnetic flux surface. 
Shape of the poloidal field line / flux surface
as function $z(x)$ for the solutions S4 (and S4b, S3, S1) 
and S9 (and S3c2). 
}
\end{figure}

\begin{table*}[ht]
\caption[]{
Comparison of leading parameters for the wind solution.
Magnetic flux distribution $\tilde{\Phi}/{\tilde{\Phi}}_{\star}$,
shape of the poloidal field line $ z(x)$,
iso-rotation parameter $\omf $,
sound speed at the injection radius $\css$,
magnetization at the injection radius $\sigs $,
cylindrical Alfv\'en radius $\xa $,
cylindrical injection radius $\xss$,
total energy $E$, normalized to $m_{\rm p} c^2$
normalized total angular momentum $\tilde{L} = L/E$, 
asymptotic velocity $u_{{\rm p}\infty}$,
and angular momentum parameter of the black hole $a$.
Other Parameters are: $\Gamma = 5/3$, 
$\ups=0.006$ (S3-S9), $\ups = 0.17$ (S3q, S3u2), $\ups = 0.21$ (S3u3)}
%
\begin{flushleft}
\begin{tabular}{llcclclrrlrrr} \hline

& prescribed & & 
  calculated & & & & & & & & \\ \hline

%
%
& $\tilde{\Phi}/{\tilde{\Phi}}_{\star}$ & $ z(x)$ & $\omf $ & 
$\css$ & $\sigs $ & $\xa $ & $\xss$ & $E$ & $\tilde{L}$ & 
$u_{{\rm p}\infty}$ & $ a $ \\ 
S3 & $\sim 1 $ & $0.1(x-x_0)^{6/5}$ & 0.035 & 0.05165 & 979.4 &
    22.931 & 8.3 & 2.7887 & 20.04 &
$ 2.531 $ & $ 0.8 $ \\
S3c2 &$\sim 1$& $0.1(x-x_0)^{3/2}$ & 0.035 & 0.0529 & 1356 & 
 22.931 & 8.3 & 2.764 & 19.95 & 2.58 & $ 0.8 $ \\
S4 & $\sim x^{-1/2} $ & $0.1(x-x_0)^{6/5}$ & 0.035 & 0.049 & 2380 &
    22.931 & 8.3 & 2.7879 & 20.04 & 
 2.60 & $ 0.8 $  \\
S4b & $\sim x^{-1/2} $ & $0.1(x-x_0)^{6/5}$ & 0.014 & 0.0390 & 14680 &
    57.0 & 15.3 & 2.6730 & 47.07 & 
$ 2.48 $ & $ 0.8 $ \\
S9 &$\sim x^{-1/2}$ & $0.1(x-x_0)^{3/2}$ & 0.035 & 0.05165 & 2777 &
  22.92 & 8.3 & 2.7572 & 19.93 & 2.57 & $ 0.8 $ \\
S3q &$\sim 1 $ & $0.1(x-x_0)^{6/5}$ & 0.14 & 0.31 & 480 &
  5.83 & 3.3 & 8.917 & 6.616 & 8.48 & $ 0.8 $ \\
S3u &$\sim 1 $ & $0.1(x-x_0)^{6/5}$ & 0.14 & 0.27 & 100 &
  5.33 & 3.3 & 3.16 & 5.69 & 2.96 & $ 0.8 $ \\
S3u2 &$\sim 1 $ & $0.1(x-x_0)^{6/5}$ & 0.14 & 0.27 & 82.5 &
  5.33 & 3.3 & 4.66 & 6.35 & 4.55 & $10^{-8}$ \\
S3u3 &$\sim 1 $ & $0.1(x-x_0)^{6/5}$ & 0.14 & 0.27 & 205.7 &
  5.33 & 3.3 & 4.65 & 6.35 & 4.48 & $10^{-8}$ \\
\hline
\end{tabular}
\end{flushleft}
\end{table*}

\subsubsection{The magnetic field distribution}
%
The normalized magnetic field distribution is prescribed by
\begin{itemize}
\item {\em the shape of the field line}, $z(x)$,
\item {\em the magnetic flux distribution},
$\Phi(x)= \tilde{\Phi}(x)\,\sqrt{-g/(\rho^2 \Delta)}$.
\end{itemize}
We apply different functions for $z(x)$ and $\tilde{\Phi}(x)$ in order 
to investigate the influence of collimation, rotation and magnetic
flux distribution on the acceleration of matter. 
One example is $z(x)=0.1 (x-x_0)^{6/5}$ describing an almost conical
surface with only a slight collimation (see Fig.\,2).
Here, $x_0$ defines the intersection of the field line with the
equatorial plane, with $x_0$ somewhat smaller than $\xss$. 
The idea behind this choice is that the matter is expected to couple 
to the jet magnetic field {\em above} the accretion disk 
(with $z(\xss) > 0$).
An example for the magnetic flux distribution is 
$\tilde{\Phi}(x)= (x/\xss)^{-1/2}$, 
resulting in magnetic flux function $\Phi(x)$ decreasing with 
radius faster than a monopole where $\Phi(x)=1$.
%
%

Prescribing both the flux distribution and the shape of the flux surface
does not over-determine the problem.
The magnetic flux function $\Phi$ describes the opening of the magnetic 
flux tubes.
With $z(x)$, the {\em shape} of the flux surface chosen, the choice of
the flux {\em function} just defines the position of the ``other'' flux 
surfaces.
In a fully self-consistent approach, the field structure is determined by
the solution of the Grad-Shafranov equation. 
Such solutions are not yet available.


\subsubsection{The plasma temperature}
%
The temperature distribution along the field line follows a polytropic
gas law, $T = T_{\star} \left(n/n_{\star}\right)^{\Gamma-1}$.
In our approach the temperature at the injection radius $\xss$ is
determined by choosing the sound speed at this point, $\css$,
\begin{equation} 
T_{\star} = \frac{\Gamma -1 }{\Gamma}
\left(\frac{\css^2}{\Gamma - 1 - \css^2}\right)
\frac{m_{\rm p}c^2}{k_{\rm B}}
\end{equation}
For typical parameters applied in our calculations, 
$\css = 0.05$, $\Gamma = 5/3$ 
this gives a gas temperature of the disk corona of about 
$1.5\times10^{10}$K
at a jet injection radius $\xss = 8.3$. 
This temperature is in rough agreement with the disk temperature of the
advection dominated accretion disk models at small radii 
(Narayan et al. 1998).
A smaller $\xss $ requires a higher sound speed parameter
implying a higher temperature $T_{\star}$.

\subsubsection{The iso-rotation parameter $\omf$ }
%
The iso-rotation parameter $\omf(\Psi)$ of the field line is determined
from the position of the injection radius $\xss$.
This choice corresponds to the interpretation often applied for $\omf$
as the ``angular rotation of the field lines''.
Here, we assume that the field lines are anchored in a Keplerian
disk,
$\omf \simeq \Omega_{\rm Disk} \simeq \Omega_{\rm K}(\xss)$.
The angular velocity of the last stable circular orbit around a Kerr
black hole is
$\omf(\xss) \sim \pm (\xss^{3/2}\pm a)^{-1}$ 
(the $\pm$ stands for co-rotation or retrograde rotation, respectively).
For a radial position not too close to the black hole, the angular
velocity in the accretion disk follows its Newtonian value.
Close to a black hole $\omf $ is limited due to the `rotation of space'
$\omega$.
An injection radius $\xss=8.3$ gives $\omf=0.04$ which is about $0.1\,\omh$
for $a=0.8$.

\setlength{\unitlength}{1mm}
\begin{figure}
\parbox{60mm}
\thicklines
\epsfxsize=90mm
\epsfxsize=90mm
\epsfxsize=90mm
\caption
{Solution S3. Properties of the critical wind solution along a given
flux surface (see parameters in Tab.\,1).
The small window shows the solution branches around the slow magnetosonic 
point enlarged.
The wind branch is the one with increasing velocity.
The critical (magnetosonic) points are located at the intersections
of the two solution branches (see Appendix C for details).
{\it Top} Poloidal velocity $\alpha \up $ (in $c$).
The asymptotic jet velocity of $\up = 2.5 $ is reached after
about $x = 10^8 $.
{\it Middle} 
Normalized proper particle density $n$ 
({\it thick line}) and
temperature $T$ in K ({\it thin line}).
{\it Below} 
Normalized poloidal ({\it thick line}) and 
toroidal ({\it thin lines}) field strength, $\bp, B_{\phi}$.
Note that the injection radius is $\xss = 8.3$.
}
\end{figure}

\section{Results and discussion}
%
We now discuss our numerical solutions of the general relativistic
magnetic wind equation for different field geometries and input
parameters.
With the prescribed poloidal field our solution is uniquely defined by
the conditions along the jet foot point and the condition of regularity
across the magnetosonic points.
Due to the stationarity assumption and the prescription of the field
distribution,
the spatial range of the computation is in principle not limited in 
radius.
This is essential if one considers the huge size of Galactic superluminal
jets in terms of the size of the central object.

In general, we show that the acceleration of plasma from regions close 
to a black hole to the speed of $0.92\,c$ observed for Galactic
superluminal motion is possible to achieve.
Depending on the poloidal magnetic field distribution, the asymptotic 
speed of the jet is reached at a radius of about 100 gravitational radii.

For comparison the leading parameters for our astrophysical solutions are 
summarized in Tab.\,1.
For illustration, we show the example solution S1 demonstrating the typical
features of the wind solution branches in the case of super- or sub-critical 
parameters (Fig.\,C.1, Appendix C). 
The meaning of our figures is explained in detail in Appendix C.

\subsection{The wind solution -- a collimating relativistic jet}
%
The time scale for the superluminal \grs jet is at least one month until
the blobs become invisible in radio light.
Mirabel \& Rodriguez (1994) estimated that the ejection event for a blob
lasts about 3 days.
This time period would correspond to a value of $\omf = 0.016$
(for $M=5\,\msun$) and an injection radius of about $\xss \simeq 15$.
The orbital period of the foot points rotating at the marginally stable
orbit (for $a=0.8$) is an order of magnitude less.
The time scale derived for the intervals between the emission of jet knots
is much larger as the period of the marginally stable orbit.
The true location of the jet origin not yet known.
Therefore, we suggest that the jet foot point should be located outside
the marginally stable orbit in order to maintain a jet flow for some time.
For the first set of solutions we chose a foot point radius of $\xss = 8.3$
or $\xss = 15.3$.

The fact that the kinematic time scale of the blobs is at least 10 times
larger than the time scale for the generation of the blobs supports the
assumption of stationarity in our calculations. 
Clearly, on the long-term time evolution the presence of the blobs them self
tells us that the jet flow is time-dependent. 

Compared to the other solutions in this sample with $\xss = 8.3$, 
solution S3 is weakly magnetized (Fig.\,3).
The initial opening angle of the magnetic flux surface is large (Fig.\,2).
The magnetic flux function $\tilde{\Phi}(x)$ is constant along the
field line.
The asymptotic poloidal velocity of $\up = 2.5$ is reached beyond
a radius $x \simeq 10^8$ 
(corresponding to a distance from the black hole of 
$z(x) \simeq 4\times10^8$).

Figure 3 also shows the distribution of other dynamical variables.
The poloidal field strength $B_{\rm p}$ decreases with the opening
of the magnetic flux surfaces.
While the poloidal field distribution is prescribed in our approach,
the toroidal magnetic field profile is a result of computation
and therefore determined by the critical wind solution.
At the injection point the toroidal field strength is about two times
smaller than the poloidal component.
Outside the Alfv\'en radius the toroidal field becomes much larger
than the poloidal component.
For large radii the magnetic field helix is dominated by the toroidal
component.
In this region we find the toroidal field component following a power law 
distribution $d(\log B_{\phi}) \simeq d(\log x)$.
Therefore, in the asymptotic part the poloidal electric current is almost
constant $I \sim xB_{\phi} \sim {\rm const.}$
In relativistic MHD electric |fields cannot be neglected.
The electric field orientation is perpendicular to the magnetic
flux surfaces and the field strength is $|\vec{E}_{\perp}| = (R/R_L) B_P$.
Therefore, the electric field is dominating the poloidal magnetic field
outside the light cylinder.

Density and temperature are interrelated by the polytropic gas law.
At the injection point the gas temperature $T\simeq 10^{10}$K (Fig.\,3).
The proper particle density at the injection point $\nss$ depends from
the choice of the mass flux (in units of the magnetic flux).
Therefore, the calculated density profile $n(x)$ may be applied to different
mass flow rates (as long as the magnetization $\sigs$ is the same).
Density and temperature decrease rapidly along the field line following the
polytropic expansion.
For $x$ \gax 30 the proper particle density follows a power law 
$n/\nss = 4\times 10^{-5} x^{-1.8}$.
%
%
At $x\simeq 1000$ the gas temperature is about $10^6$K.
Therefore we can estimate the size of a X-ray emitting region 
of about several $1000\,\rg$ in diameter.
For the example of \grs this corresponds to $3.5\times10^{-9}$ arcsec.
It would be interesting to calculate the X-ray spectra of such an
relativistically expanding high temperature gas distribution.

Solution S3c2 has the same distribution of the magnetic flux 
function $\Phi$ as solution S3.
The magnetic flux surfaces, however, are collimating more rapidly.
The derived critical wind solution has a higher magnetization,
although the terminal speed and the total energy density $E(\Psi)$ 
of the S3c2 solution is similar to S3.
Because of the higher magnetization type S3c2 jet solutions have a
correspondingly lower mass flow rate.
The asymptotic speed is reached already at about $x=1000$
equivalent to a distance from the central black hole of about $z=3200$.

Solution S9 relies on the same magnetic flux surface as S3c2.
As a difference to S3c2, the magnetic flux function decreases with radius
implying a (spatially) faster magnetic field decay.
As a consequence, the jet reaches its asymptotic velocity of $\up = 2.57$ 
even at about $x=100$.
The derived flow magnetization is higher compared to S3c2 and S3 balancing
the fast decay of the magnetic field distribution and we obtain the same
asymptotic speed.
This is interesting because it proves that not only the magnetization,
but also the distribution of the magnetic flux along the field line
determines the asymptotic speed.

Note that solution S9 reaches the same asymptotic speed as S3c2
only because of its higher magnetization.
Indeed, a solution similar to S9, but having the same magnetization 
$\sigs = 1356$ as for S3c2, only reaches an asymptotic speed of
$\up = 1.81$ (not shown).
Also, such a solution would be only very weakly magnetized in the
asymptotic regime 
as the normalized flow magnetization changes as
$\sigma \sim 1/\sqrt{x}$ for the $\Phi \sim 1$ solutions
or $\sigma \sim 1/x$ for the $\Phi \sim 1\sqrt{x}$ solutions
\footnote{
However, 
in the hot wind equation it is not possible to change only one single
parameter in order to obtain a new set of critical wind solutions.
In the case discussed above,
with the decreased magnetization (i.e. an increased mass flow rate),
the Alfv\'en radius is correspondingly smaller 
(here, $\xa = 21.11$ compared to $\xa = 22.93$ for S9).
},
respectively.
Similarly, in comparison, the asymptotical toroidal magnetic field is
weaker by some orders of magnitude (a factor ten at $x=1000$).
In all the solutions presented in this paper the {\em asymptotic} jet
is dominated by the kinetic energy.
For the solutions with the large injection radius $\xss = 8.3$,
the magnetic energy is being converted into kinetic energy almost
completely already at a radius of about several 100 gravitational 
radii.

Solution S4 has the same magnetic flux distribution as S9, however,
the field line is only weakly collimating.
The asymptotic jet speed and the magnetization parameter is about the
same.
Only, the initial acceleration is weaker because the magneto-centrifugal
mechanism works less efficient in the field with a smaller opening angle.

%
Solution S4b has essentially the same field distribution as S4,
but the injection radius is chosen larger.
Therefore, the iso-rotation parameter $\omf$ is decreased by a factor of
$(8.3/15.3)^{3/2}$. 
As a result, a critical wind solution with a comparable asymptotic speed
could be obtained only for a very high plasma magnetization.
This proves that highly relativistic jets can be expected even if the
jet is not emerging from a region close to the black hole.
Such a solution is feasible if the mass flow rate in the jet decreases
with radius faster than the field strength (or flux distribution).
The question remains whether such field strengths can be found at this
position.

We summarize the results of this section.
The asymptotic speed is determined by the plasma magnetization and
the distribution of the magnetic flux along the field line.
The shape of the magnetic flux surface determines the velocity
{\em profile} along the field, 
thus, the position where the asymptotic velocity is reached.
Highly relativistic outflows can be obtained even if the jet foot
point is {\em  not} very close to the black hole. 
However, in this case a high plasma magnetization is necessary.
But this seems to be in contradiction to the accretion
disk theory (see below).

\setlength{\unitlength}{1mm}
\begin{figure*}
\parbox{60mm}
\thicklines
\epsfxsize=90mm
\epsfxsize=90mm
\epsfxsize=90mm
\epsfxsize=90mm
\caption
{Wind solutions 
S9 ({\it upper left}),
S3c2 ({\it lower left}),
S4 ({\it upper right}), 
S4b ({\it lower right}). 
Branches of poloidal velocity $\alpha \up $ along the field line in
units of the speed of light. For the solution parameters see Tab.\,1.
See caption of Fig.\,3 for further explanation.
}
\end{figure*}

\subsection{The role of the magnetization}
%
The magnetic acceleration of jets and winds can be understood
either as a consequence of converting Poynting flux (magnetic energy)
to kinetic energy
or due to Lorentz forces along the poloidal field line.
In general, the higher the plasma magnetization the more energy can be
transformed into kinetic energy of the wind.
It has been shown theoretically for a cold wind
that the relation between
magnetization and asymptotic velocity is that of a power law,
$u_{{\rm P}\infty} \sim {\sigma_{\rm M}}^{1/3}$,
for conical outflows (Michel 1969)
and for collimating flows (Fendt \& Camenzind 1996),
if the flux distribution is the same, respectively.
However, both papers do not consider gravity 
(and no general relativistic effects).
The new solutions presented in this paper are in general agreement
with those results
in the sense that a higher magnetization leads to a higher velocity.
However, we are dealing with the {\em hot} wind equation and cannot derive
a power law distribution from Tab.\,1,
since the other wind parameters may vary between the different solutions.
In difference to the cold wind solutions the magnetization is not a free
parameter.
Instead, it is fixed by the regularity condition at the magnetosonic
points.

The wind magnetization is determined by the disk properties at the jet
injection points along the disk surface.
For a standard thin disk model that the ratio of the mass flow rate in
the jet to the disk accretion rate is about 1\% (Ferreira 1997).
The observational data for various jet-disk systems are consistent with
this theoretical value.
The accretion disk magnetic flux can be estimated assuming equipartition
between magnetic field pressure (energy) and gas pressure (thermal energy)
in the disk (see Sect.\,3.4.1).
From Eq.\,(7) we find an equipartition field strength of about 
$ B_{\rm eq} \simeq 5\times10^8$G,
if $\alpha_{\rm vis} \simeq 0.1$ and $\rss = 10\rg$.
Equation (8) then defines an upper limit for the plasma magnetization at
the injection radius, $\sigs = 5\times 10^4 $, for 
$\dot{M}_{\rm jet}\simeq 0.1\dot{M}_{\rm acc}$.
Such a value is in general agreement with our solutions (Tab.\,1).
The maximum equipartition field strength estimated with the above given
formulae can be much larger for Galactic black hole jet sources as for
AGN (see Eq.\,7).
For a low black hole mass (with a smaller horizon) the disk comes closer
to the singularity and therefore becomes hotter.

Again, we note that our estimate for the magnetization comes from
comparison of different disk models (Sect.\,3.4.1.).
However, this does not mean that we {\em apply} a certain disk model
for our computations.


Finally, we come back to the wind solutions S4 and S4b.
As already mentioned, these solutions demonstrate that the jet origin 
must not be necessarily close to the black hole.
One may think that a strong magnetization at larger disk radii would 
do the job.
On the other hand, the equipartition field strength in the disk 
decreases with radius 
implying that the highest magnetization and, thus, jet velocities must
be expected from the inner part of the disk.
Only, if the mass transfer rate from the disk into the jet decreases 
more rapidly with radius than the field strength, 
the {\em magnetization} increases.

\subsection{The influence of the rotating black hole}
%
As the main issue of our paper is the search for MHD wind solutions
in Kerr metric,
it is necessary to clarify the role of general relativity for the jet
acceleration itself.
Clearly, at an injection radius of $\rss = 8.3$ general relativistic
effects are not very dominant.

For comparison we have calculated wind solutions for a smaller
injection radius $\xss = 3.3$ (solution S3q, S3u, Fig.\,D1, D2).
The main effect is a much higher asymptotic velocity resulting from
the rapid rotation, $\omf$, at the smaller radius $\xss$.
With our choice $\omf=0.14$ the asymptotic velocity drastically increases
from $\up=2.503$ (S3) to $\up = 8.4792$ (S3q).
In order to obtain the critical solution for the higher rotation rate, 
the wind parameters have to be changed accordingly.
$\sigs$ is decreased by a factor of two, while $\css$ and $\ups$
must be increased substantially.
The large sound speed is in agreement with the smaller injection radius,
since a higher disk temperature and pressure is expected close to the hole.
The Alfv\'en radius is decreased by a factor of four,
however, its location relative to the outer light cylinder remains the
same.

The limiting case of Minkowski metric can be achieved by setting
$M=0$ and $a=0$ in the Boyer-Lindquist parameters (see Appendix A).
For such a wind solution (S3u2)
the magnetization is lower, although the asymptotic wind speed is
the same as in the Schwarzschild case (see Fig.\,D2).
This becomes clear if we take into account that for S3u2 the wind flow
does not have to overcome the gravitational potential
Thus, less magnetic energy is needed to obtain the same asymptotic
speed by magnetic acceleration.
Further, we find from solutions for different angular momentum
parameters $a$ that in general the wind flow originating from a black
hole with a smaller $a$ is faster.
As an extreme example we show the solution S3u3 calculated with $a\simeq0$ 
but otherwise the same parameter set (see Fig.\,D2).
This solution is magnetized stronger compared to the case of $a=0.8$,
thus, resulting in a higher asymptotic wind velocity.
We believe that the reason for such a behavior is the fact that the
effective potential of a black hole weakens (at this location) for 
increasing values of $a$.
Therefore, less magnetic energy is necessary to overcome the effective
potential.

In the end, the results of this section are not surprising.
They demonstrate that the wind/jet is basically {\em magnetically driven}.
As a consequence, the acceleration takes place predominantly across
the Alfv\'en point as expected from MHD theory.
Therefore, the scenario is similar to the case of classical pulsar theory
in Minkowski metric.
For relativistic jets with a high magnetization the Alfv\'en point is
always very close to the light surface, which is defined by the 
angular velocity of the field line foot point.
Usually, the Alfv\'en point is located at a radius large compared to
the gravitational radius.
Thus, the influence of the general relativistic metric is marginal.
Only, if the Alfv\'en radius comes close to the hole, the choice of
the metric will determine the jet acceleration.

\subsection{The question of collimation}
%
The huge size observed for the knots of the Galactic superluminal
sources 
leaves the possibility that the jet is basically {\em un-collimated}.

Our numerical solutions have shown that the asymptotic speed of the
jet does not depend very much on the degree of collimation in the flow.
That speed is reached within a distance of about $ 10^8\rg$.
However, the observed upper limit for the knot size is still a factor 
10 larger.
Therefore, from our solutions, the observed knots are consistent with
both a collimated and an un-collimated jet flow.
In particular, solution S9 which is more collimated, has the same 
asymptotic speed as solution S4.

In the case of extragalactic jets a high degree of collimation is 
indicated.
The ``lighthouse model'' by Camenzind \& Krockenberger (1992) gives
opening angles of only 0\fdg1 for the quasar 2C273 or 
0\fdg05 for typical BL\,Lac objects.
The question arises 
whether there could be an intrinsic difference 
between the jets of AGN and Galactic high energy sources.
Why should Galactic superluminal jets be un-collimated? 
A difference in the jet magnetization seems to be unlikely since the jet
velocities are comparable.
We hypothesize that if the jets of these sources are systematically 
different,
this should rather be caused by the conditions in the jet environment.
If the jets are collimated by external pressure, 
a different external/internal pressure ratio will affect the degree of
jet collimation.
Extragalactic jets are believed to be confined by 
an external medium
(see Fabian \& Rees 1995, Ferrari et al. 1996).
It is likely that Galactic superluminal sources provide an example where 
the jet pressure exceeds the pressure of the ambient medium.
While AGN jets bore a funnel through the galactic bulge,
Galactic superluminal jets freely expand into the empty space.
Such a picture seems to be supported by the fact that the Galactic
superluminal jet knots move with constant velocity over a
long distance.

%
\section{Summary}
%
We have investigated magnetically driven superluminal jets originating
from a region close to a rotating black hole.
The stationary, general relativistic, magnetohydrodynamic wind equation
along collimating magnetic flux surfaces was solved numerically.
The wind solutions were normalized to parameters typical for Galactic 
superluminal sources.

The assumption of stationarity allows us to calculate the jet velocity
on a {\em global} scale over a huge radial range in terms of radius of
the central source.
The wind is launched close to the rotating black hole at several
gravitational radii.
The calculation was performed up to a radius of $10^4$ gravitational radii,
but is in general not limited in radius.
In some cases the asymptotic speed may be reached only at a distance of
several $10^8$ gravitational radii.
Different magnetic field geometries were investigated.
The model allows for a choice of the {\em shape} of the magnetic
flux surface {\em and} the {\em flux distribution} of that field.

The physical wind solution is defined by the regularity condition
at the magnetosonic points.
As the poloidal field is prescribed, the choice of the following 
input parameters determines the wind solution completely,
(i) the injection radius of the matter into the jet, 
(ii) the injection velocity and
(iii) the plasma magnetization (the ratio of magnetic flux to mass flux).
The results of our numerical computation are the following.

\begin{itemize}

\item
In general, the observed speed for Galactic superluminal sources of more 
than 0.9\,c can be achieved.

\item
The flow acceleration is magnetohydrodynamic and takes place predominantly
around the Alfv\'en point.
General relativistic effects are important only if the wind originates
very close to the black hole.
In order to overcome the gravitational potential, the critical wind
solution must be higher magnetized in order to reach a similar 
asymptotic speed.
This has been proven by calculating the Schwarzschild and Minkowski limit 
of the wind equation.

\item
For a fixed magnetic field distribution the asymptotic jet velocity depends
mainly on the plasma magnetization,
in agreement with earlier papers (Michel 1969, Fendt \& Camenzind 1996).
The higher the plasma magnetization, the higher the final speed.
The velocity distribution along the magnetic field shows a saturating
profile depending on the distribution of the magnetic flux.

\item
The magnetic flux distribution along the field line also influences the
plasma acceleration.
Since the real field distribution is not known, we have considered
two cases which show the typical behavior and which are probably
close to the reality.
We find that the jet velocity in a (spatially) faster decaying field 
can be the same as long as the magnetization at the injection
point is high enough in order to balance the effect of the decrease in
field strength.

\item
For jet solutions {\em not} emerging from a region close to the
black hole, a highly relativistic velocity can be obtained if the flow
magnetization is sufficiently large.
However, one we hypothesize that the field strength required for
such a magnetization can be generated only close to the black hole.

\item
Investigation of flux surfaces with a different degree of collimation
has shown that both field distributions allow for a relativistic velocity.
The asymptotic jet velocity is reached considerably earlier in the case of
the faster collimating flux surface.
The jet reaches its asymptotic speed at a distance from the injection
point of $3000\rg$ or $10^5\rg$, depending on the degree of collimation.
The latter we measure with the opening angle of the collimating flux
surface at this point and is about 15\degr or 45\degr, respectively.
This distance is below the observational resolution by several orders of
magnitude.
Therefore, the question of the degree of collimation for the
superluminal jets of \grs and \gro could not be answered.

\item
Motivated by the huge size of the observed knots in the Galactic
superluminal jets, we point out the possibility that the jet collimation
process in these sources may be intrinsically different in comparison
to the AGN.
For example, the upper limit for the knot diameter in \grs
is about $10^9$ Schwarzschild radii, 
which is distinct from typical estimates for AGN jets with diameters of
about 100 - 1000 Schwarzschild radii.

\item
The gas temperature at the injection point is about $10^{10}$K which is
one order larger than the disk temperature at this point.
With the polytropic expansion the temperature decreases rapidly to about
$10^6$K at a distance of $5000$ Schwarzschild radii from the source.
Both the temperature and the mass density follow a power law distribution
with the radius.

\item
The calculations show that the jet magnetic field is dominated by the
toroidal component.
Similarly, the velocity field is dominated by the poloidal component.

\end{itemize}

In summary, our numerical calculations have shown that the highly
relativistic speed observed for galactic superluminal sources can be
achieved by magnetic acceleration.
For a given magnetic flux surface we obtain the complete set of physical
parameters for the jet flow.
The calculated temperature, density and velocity profile along the jet
would provide a interesting set of input parameters for computing the
spectral energy distribution.

 
%
\begin{acknowledgements}
C.F. thanks Mikael Jensen at Lund Observatory for discussions and the
analytical proof of the equations in Appendix B.
We acknowledge fruitful comments by the referee Dr. Koide helping
to improve the presentation of this paper, especially Sect.\,4.3.
\end{acknowledgements}

\appendix
%
\section{Parameters of Kerr metric}
%
For the reason of completeness, here we list the parameters 
applied in the equations of Kerr geometry.
In Boyer-Lindquist coordinates with the parameters
\begin{eqnarray} 
\rho^2   & \equiv & r^2 + a^2\,\cos^2\theta,
                                      \quad\quad\quad\quad\quad\quad
\Delta  \equiv  r^2 + a^2 - 2M\,r,                     \nonumber\\
\Sigma^2 & \equiv & (r^2 + a^2)^2 -a^2\Delta\sin^2\theta,
					    \quad\quad 
\tom  \equiv (\Sigma/\rho)\,\sin\theta,                \nonumber\\
\omega & \equiv & 2\,a\,M\,r / c\,\Sigma^2, 
                             \quad\quad\quad\quad\quad\quad
\ \ \alpha  \equiv  \rho\,{\sqrt{\Delta}} / \Sigma,        \nonumber
\end{eqnarray}
the components of the metric tensor are defined as
\begin{eqnarray}
g_{00} & = &  \sm ( 2 r/\rho(r,\theta)^2 - 1 )\nonumber \\
g_{03} & = & - \sm  2 r a \sin(\theta)^2 / \rho(r,\theta)^2\nonumber \\
g_{11} & = &  \sm \rho(r,\theta)^2 / \Delta(r,\theta)\nonumber \\
g_{22} & = &  \sm \rho(r,\theta)^2 \nonumber \\
g_{33} & = & \sm\Sigma(r,\theta)^2\sin(\theta)^2 / \rho(r,\theta)^2\nonumber \\
g & \equiv & {\rm Det(g_{\mu \nu})} 
          = - g_{11} g_{22} ( g_{30}^2 - g_{00}g_{33}) 
       \nonumber 
\end{eqnarray}
In our paper we have chosen a negative sign of the metric, $\sm = -1$.

\section{Wind polynomial}
Here we provide the polynomial coefficients of the
general relativistic magnetohydrodynamic wind equation Eq.\,(4).
For a derivation, see Camenzind (1986), Takahashi et al. (1990), or
Jensen (1997).
The specific angular momentum, properly normalized, is  
\begin{equation}
\aL = - (g_{03} + \omf g_{33}) / (g_{00} + \omf g_{03})
\end{equation}
For convenience we define the following parameters,
\begin{eqnarray}
C_1 & = & \frac{\css^2}{\Gamma-1-\css^2} \left(\ups \sqrt{\frac{g_{\star}}{g}} 
       \frac{\Phi}{\phs}\right)^{\Gamma-1} \!\!\!\!\!, \quad
C_2 = \sqrt{-g} \frac{\phs}{\Phi \sigs}  \nonumber\\
D_1 & = & g_{00}+2 \omf g_{03}+\omf^2 g_{33}, \quad
D_2 = (1 - \omf\aL)^2 \nonumber \\
D_3 & = & -(g_{33} +2 \aL g_{03} +\aL^2 g_{00})/(g_{03}^2-g_{00} g_{33})
      \nonumber
\end{eqnarray}
With the corresponding values at the injection radius $\xss$
the total specific energy density of the flow $E$ is defined as
\begin{equation}
E^2 =  \frac{-\sm \mu_{\star}^2 (\ups^2+1) (D_{1\star}+\sm M_{\star}^2)^2}
       { (D_{1\star}+2 \sm M_{\star}^2) D_{2\star} + D_{3\star} M_{\star}^4} ,
\end{equation}
where $M_{\star}$ denotes the Alfv\'en Mach number at the injection radius.
The polynomial coefficients of the wind equation (4) are 
\begin{eqnarray}
\tilde{a}_{1,2n+2m} & = & C_2^2\nonumber  \\
\tilde{a}_{2,2n+m}  & = & 2 \sm C_2 D_1\nonumber \\
\tilde{a}_{3,2n}    & = & D_1^2+C_2^2+\sm E^2 C_2^2 D_3 \nonumber \\
\tilde{a}_{4,2n-m}  & = & 2 \sm C_2 D_1+2 E^2 C_2 D_2\nonumber \\
\tilde{a}_{5,2n-2m} & = & D_1^2+\sm E^2 D_1 D_2\nonumber \\
\tilde{a}_{6,n+3m}  & = & 4 C_1 C_2^2\nonumber \\
\tilde{a}_{7,n+2m}  & = & 6 \sm C_1 C_2  D_1\nonumber \\
\tilde{a}_{8,n+m}   & = & 2 C_1 D_1^2+4 C_1 C_2^2+\sm 2 E^2 C_1 C_2^2 D_3
\nonumber \\
\tilde{a}_{9,n}     & = & \sm 6 C_1 C_2 D_1+2 E^2 C_1 C_2 D_2 \nonumber \\
\tilde{a}_{10,n-m}  & = & 2 C_1 D_1^2\nonumber \\
\tilde{a}_{11,4m}   & = & 6 C_1^2 C_2^2\nonumber \\
\tilde{a}_{12,3m}   & = & 6 \sm C_1^2 C_2 D_1\nonumber \\
\tilde{a}_{13,2m}   & = & C_1^2 D_1^2+6 C_1^2 C_2^2+\sm E^2 C_1^2 C_2^2 D_3
\nonumber \\
\tilde{a}_{14,m}    & = & 6 \sm C_1^2 C_2 D_1\nonumber \\
\tilde{a}_{15,0}    & = & C_1^2 D_1^2\nonumber \\
\tilde{a}_{16,5m-n} & = & 4 C_1^3 C_2^2\nonumber \\
\tilde{a}_{17,4m-n} & = & 2\sm C_1^3 C_2 D_1\nonumber \\
\tilde{a}_{18,3m-n} & = & 4 C_1^3 C_2^2\nonumber \\
\tilde{a}_{19,2m-n} & = & 2\sm C_1^3 C_2 D_1\nonumber \\
\tilde{a}_{20,6m-2n}& = & C_1^4 C_2^2\nonumber \\
\tilde{a}_{21,4m-2n}& = & C_1^4 C_2^2\nonumber 
\end{eqnarray}
All coefficients with the same second index have to be summed up,
$A_i = \sum_{j} \tilde{a}_{j,i}$.
The polytropic indices $n=5$, $m=3$ give a polynomial of 16th order.

\section{Example wind solution in Kerr metric}
%
Here we show an example solution of the wind equation (4).
The parameters are chosen such that a variation of $\sigs$ and
$\css$ clearly demonstrates the criticality of the wind solution.
They do not necessarily match astrophysical constraints.
However, the asymptotic poloidal velocity is comparable to the speed of
the Galactic superluminal sources.
The solution (solution S1) considers a highly magnetized plasma flow 
with $\sigs \simeq 5\,10^4 $.
The flux geometry is that of a slightly collimating cone with an opening
angle decreasing with distance from the source.

Figure C1 shows the solution branches with a positive $\up^2$.
An overlay of solutions for three parameter sets is displayed in order to
show the typical behavior of wind solution.
There is only one unique solution, the critical solution, with one branch
continuing from small to large radii without any gaps in $\up$ or $x$.
The magnetosonic points are located at the intersections of the solution
branches of the critical solution.
The {\em critical} wind solution is regular at all three magnetosonic points.
It is defined by a {\em unique} set of the parameters $E,L$ and $\sigma$ 
(for $\omf$ prescribed).
In the critical solution the slow magnetosonic point is passed close to
the foot point of the jet.
The Alfv\'en point is located at $x = 31$ and the fast magnetosonic point 
not far beyond.
The asymptotic speed of the flow is $\up = 2.28$, equivalent to
$\vp \simeq = 0.9\,c$ (not shown in the Figure).

Sub- or super-critical wind parameters lead to solution branches which
are not defined for all radii or all velocities.
Even for a slight variation of these parameters the solution will be not
continuous anymore,
implying ``jumps'' or ``shocks'' across the gaps in the solution
branches.
At these locations the stationary character of the solution most probably
breaks down. 
Such solution branches are inconsistent with the assumptions and are
therefore referred to as {\em unphysical}.

\setlength{\unitlength}{1mm}
\begin{figure}
\parbox{60mm}
\thicklines
\epsfxsize=90mm
\caption
{Example solution S1.
Overlay of solutions $\up(x)$ for three different parameter sets.
$\sigs = 49830$, $ \css = 0.4585$ gives the critical solution
which is regular across the magnetosonic points.
The critical {\em wind} solution is the continuous branch starting
with low velocity and accelerating to high speed.
The magnetization $\sigs$ is the critical parameter for the FM point, 
whereas $\css$ is the critical parameter for the SM point.
Sub- or super-critical solutions are obtained by variation of the
parameters $\sigs $, $\css $.
The choice of 
$\sigs = 51830$, $ \css = 0.4485$ results in gaps in $x(\up)$,
the choice of 
$\sigs = 48830$, $ \css = 0.4685$ in gaps in $\up(x)$.
The other parameters are  $\xa = 31.2$, $\ups = 0.01$, $\xss = 3.0$, 
$\omf = 0.1\,\omh = 0.025$, $a = 0.8$.
}
\end{figure}

\section{The wind solution for a small injection radius}
%
For comparison, we show solutions of the wind equation with a small
injection radius $\xss = 3.3$ as well as solutions in the limit of
Minkowski and Schwarzschild metric (for a discussion see Sect.\,4.3).
Solution S3q corresponds to solution S3, however, with a magnetization
smaller by a factor of two. The asymptotic speed is $\up = 8.48$
and much larger than for S3.
Also solutions S3u, S3u2, S3u3 correspond to S3 and S3q.
However, in this case the Alfv\'en radius and the derived magnetization
parameter are lower resulting in a lower asymptotic speed.
Solution S3u is the Kerr solution for $a=0.8$, S3u3 the Schwarzschild
solution ($a=10^{-8}$), and S3u2 the Minkowski solution where we set 
$a=10^{-8}$ and $M=0$ in the Boyer-Lindquist parameters (see Appendix A).
For a comparison of all solutions see Tab.\,1.

\setlength{\unitlength}{1mm}
\begin{figure}
\parbox{60mm}
\thicklines
\epsfxsize=90mm
\caption
{Example solutions with a small injection radius $\xss = 3.3$.
Rotation rate $\omf = 0.14$.
Solution S3q with $a=0.8$, $\xa = 5.83$, $\sigs = 480$
has a high asymptotic velocity $\up = 8.48$. 
}
\end{figure}

\setlength{\unitlength}{1mm}
\begin{figure}
\parbox{60mm}
\thicklines
\epsfxsize=90mm
\epsfxsize=90mm
\epsfxsize=90mm
\caption
{Example solutions with a small injection radius $\xss = 3.3$.
Rotation rate $\omf = 0.14$.
{\it Top} Solution S3u with $a=0.8$, $\xa = 5.33$, and a smaller
$\sigs = 100$ and a lower asymptotic velocity $\up = 2.96$.
{\it Middle} Solution S3u3 in Schwarzschild metric, $a=10^{-8}$, 
$\xa = 5.33$.The asymptotic speed is $\up = 4.48$ with $\sigs = 205.7$.
{\it Bottom} Solution S3u2 in Minkowski metric, $a=10^{-8}$, $M=0$.
The asymptotic speed is $\up = 4.55$ while $\sigs = 82.5$.
}
\end{figure}


\end{document}